\documentclass[review]{elsarticle}
\usepackage[table]{xcolor}
\usepackage{lineno}

\journal{Journal of Computer Speech and Language}

\begin{document}

\begin{frontmatter}

\title{Voice Spoofing Detection Corpus for Single and Multi-order Audio Replays}
\tnotetext[mytitlenote]{This paper is submitted on 26 August, 2019.}

\author{Roland Baumann\textsuperscript{a},Khalid Mahmood Malik\textsuperscript{a*},Ali Javed\textsuperscript{a},Andersen Ball\textsuperscript{a}, Brandon Kujawa\textsuperscript{a},Hafiz Malik\textsuperscript{b}}
\address{\textsuperscript{a} Computer Science and Engineering Department, Oakland University, Rochester, MI, USA}
\address{\textsuperscript{a} Electrical and Computer Engineering, University of Michigan-Dearborn, MI, USA}

\author{}
\ead[url]{https://oakland.edu/secs/directory/mahmood}
%\author[mysecondaryaddress]{\corref{mycorrespondingauthor}}
\cortext[mycorrespondingauthor]{Corresponding author}
\ead{mahmood@oakland.edu}

%\address[mymainaddress]{^b^c^d^e318 Meadow Brook Rd, Rochester, MI 48309, USA}
%\address[mysecondaryaddress]{^f4901 Evergreen Rd, Dearborn, MI 48128}

\begin{abstract}
The evolution of modern voice-controlled devices (VCDs) has revolutionized the Internet of Things (IoT), and resulted in increased realization of smart homes, personalization and home automation through voice commands. These VCDs can be exploited in IoT driven environment to generate various spoofing attacks including the chain of replay attacks (multi-order replay attacks). Existing datasets like ASVspoof and ReMASC contain only the first-order replay recordings, therefore, they cannot offer evaluation of the anti-spoofing algorithms capable of detecting the multi-order replay attacks. Additionally, these datasets do not capture the characteristics of microphone arrays, which is an important characteristic of modern VCDs. Therefore, there exists an urgent need to have a diverse replay spoofing detection corpus that consists of multi-order replay recordings against the bonafide voice samples. This paper presents a novel voice spoofing detection corpus (VSDC) to evaluate the performance of multi-order replay anti-spoofing methods. The proposed VSDC consists of first-order- and second-order-replay samples against the bonafide audio recordings. We ensured to create a diverse replay spoofing detection corpus in terms of environment, recording and playback devices, human speakers, configurations, replay scenarios, etc. More specifically, we used 22 microphones, 25 different recording configurations, 54 different playback configurations for first-order- and second-order-replays to generate a total of 11,772 samples belonging to fifteen human speakers. Additionally, the proposed VSDC can also be used to evaluate the performance of speaker verification systems. To the best of our knowledge, this is the first publicly available replay spoofing detection corpus comprising of first-order- and second-order-replay samples. 
\end{abstract}


\end{frontmatter}

\section{Introduction}
 The growing trend of personalization, realization of smart homes, and the desire for easy control of home devices are driving factors for the tremendous evolution of Internet of Things (IoT) devices. Voice assistant, a software component of voice-controlled devices (VCD) such as Google home, Amazon Echo and Siri etc., is becoming an essential component of Internet of Things (IoT). In near future, VCDs are expected to have capability of processing both audio and video, and is expected to introduce a new subfield of IoT, called Multimedia of Thing (MoT). MoT devices, a subset of IoT devices, are equipped with microphones, cameras and speakers. VCDs are susceptible to various audio spoofing attacks such as replay attacks, voice cloning attacks, etc., while MoT devices face various multimedia spoofing challenges including deepfakes \cite{19}. 
\\ The voice assistants have enabled enormous connectivity among VCDs and are opening vistas of new research \cite{5}. Particularly, the addition of microphones arrays and speakers enable these devices to engage in two-way communication, allowing them to play audio and accept voice commands from other IoT devices. The most recognizable feature of VCDs has been the capability to connect all household IoT devices together with voice commands. Voice assistants are now being directly integrated into thermostats, refrigerators, light switches, entertainment systems and cars. It is important to mention that many IoT devices in smart homes are controlled remotely through VCDs. In addition to controlling IoT devices in the home, integrated voice assistants are also being used to bring a variety of Internet based applications related to these VCDs such as entertainment, communication, shopping, healthcare, business, banking services, etc. With so many devices in a home being able to provide a voice assistant, the system resembles the dream of science fiction television where an omnipresent computer is constantly ready to provide quick and easy verbal access. The VCDs themselves could be used to replay audio to each other forming the basis of multi-hop scenarios. The open space inside a home becomes a transmission medium through which one VCD can replay voice commands to another VCD. Figure 1 illustrates how a smart home can have many VCDs capable of speaking to each other. 
\begin{figure}
  \includegraphics[width=1\textwidth]{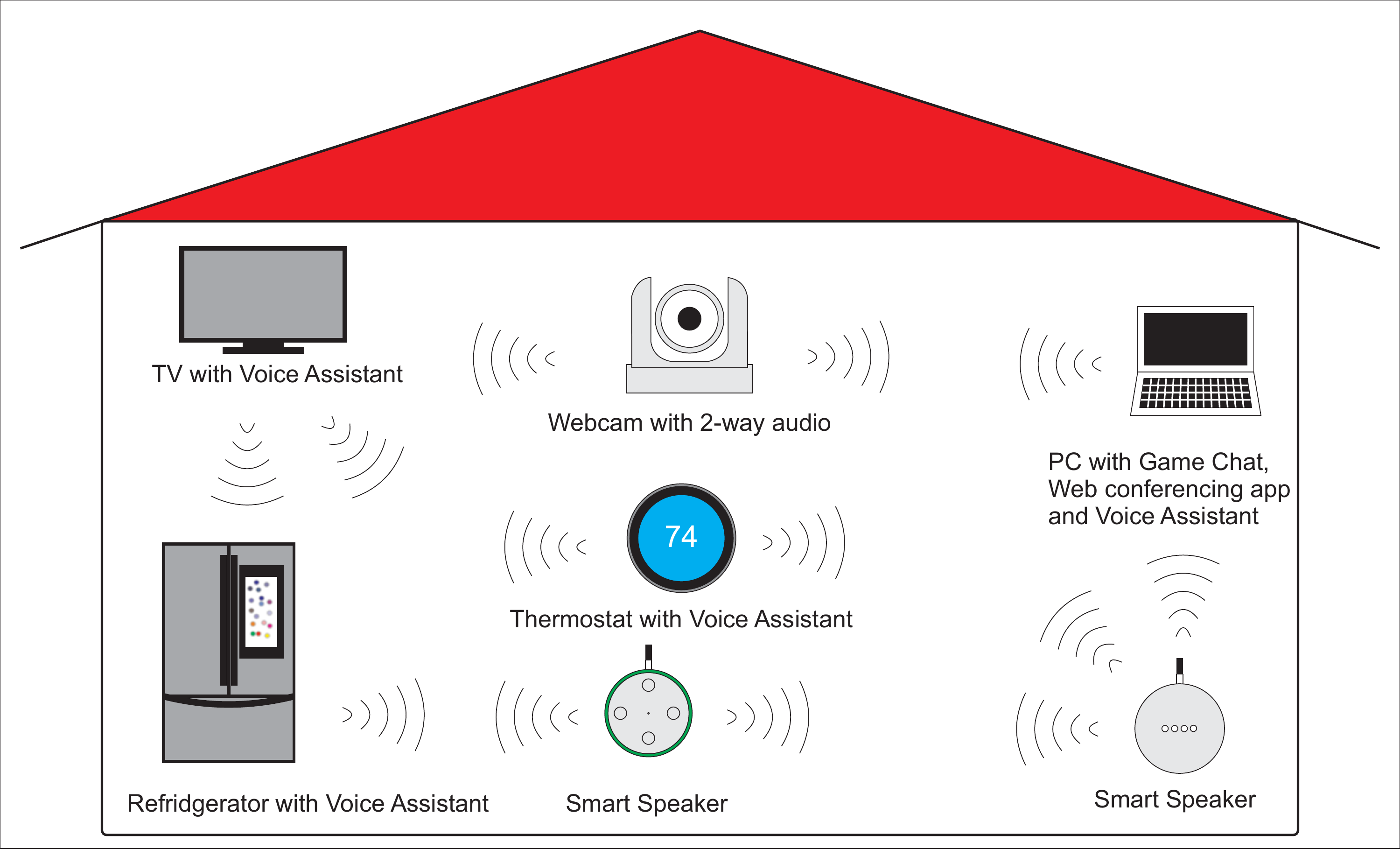}
\caption{Voice-controlled devices connectivity to send/receive audios in the home.}
\label{fig:1}       % Give a unique label
\end{figure}
\\ Most VCDs are equipped with array microphones which means they have more than one microphone. The Amazon Echo Dot 3 uses an array of 4 microphones. This array of microphones allows the VCD to determine the location of the human speaker, selection of the best microphone and use the other microphones to reject background noise. This configuration allows VCDs to pick up voice command at long distances (few meters) in less than ideal conditions. This fact enables the VCD to be more susceptible to replay attacks.
\\ Audio-specific spoofing attacks can be categorized into replay \cite{12}, speech-synthesis (SS) \cite{13}, voice conversion (VC) \cite{14} and impersonation \cite{17}. Among all audio spoofing attacks, replay attacks could be more prevalent in the future, as less tech savvy intruders can generate them and disrupt the automatic speaker verification system of a VCD based system \cite{15}. Existing spoofing datasets \cite{2,3,16} are designed for evaluation of testbeds that consider replay spoofing as a two-class classification problem. The application focus of these datasets is mainly evaluating voice driven banking systems and they only address the scenario of a one-time replay. However, we have demonstrated through the experimentation in our earlier work \cite{5} that VCDs are very vulnerable to even second-order replay attacks and are unable to clearly classify between the original and spoof samples in multi-hop scenarios. This vulnerability of VCDs can easily be exposed by an intruder to cause severe financial loss and data theft. Additionally, existing datasets i.e. ASVspoof do not contains the audio samples recorded from devices having array of microphones. Therefore, there exists a need to create a replay spoofing dataset to evaluate applications and testbeds that may involve multi-hop voice propagation scenarios and samples recorded with devices having microphone arrays. For this purpose, we designed a novel voice spoofing detection corpus (VSDC) for multi-hop replay scenarios that consist of bonafide, first-order- and second-order-replay audio samples. Additionally, we tried to ensure that our replay dataset should be diverse in terms of recording environment, background noise, recording and playback devices, microphones, speakers, replay scenarios, etc. It is important to mention that the proposed dataset is unique and first-one to consider the multi-hop replay scenarios. 
\\The main contributions of this paper are: 
\\ 1. Development of a large-scale dataset for evaluation of audio forensics testbed. 
\\ 2. Development of a multi-order replay dataset consisting of bonafide, first-order replay and second-order replay samples that can effectively be used to evaluate the performance of anti-spoofing methods in multi-hop scenarios. 
\\3. We ensured to create a diverse replay spoofing detection corpus in terms of environment, recording and playback devices, human speakers, configurations, replay scenarios, etc. More specifically, we used 22 microphones, 25 unique recording configurations, 54 unique playback configurations to generate a total of 11772 samples belonging to fifteen human speakers of different age and gender. 

\section{Landscape of Multi-hop replay attacks}
In this section we briefly discuss the landscape of replay attacks involving the voice-controlled devices (VCDs). The addition of a voice interface introduces a new attack surface to be exploited in homes, offices, businesses and hospitals. These scenarios demonstrate that multiple-replays on the VCDs can be used to exploit the systems having voice interfaces. Although we discuss the scenarios of smart homes in this paper, the threats associated with the replay attacks can go beyond the homes and voice-controlled applications being developed for smart cities, futuristic cars, and other businesses. Amazon has already launched its smart assistant Alexa for businesses automation \cite{7}. Currently, Amazon is working on healthcare apps that use smart speakers to perform various tasks \cite{6}. The details of a few representative scenarios involving the multi-order replay attacks are discussed below. It is important to mention that we have experimentally verified these scenarios.
\subsection{Scenario 1: Webcam Replay}
Shown in Figure 2 are the two scenarios where VCDs are used to replay audio to each other. In scenario 1, a compromised webcam is able to listen a user giving commands to a Google Home device. In such a scenario, a webcam can be accessed by compromising the home’s WiFi network using a tool i.e. Aircrack-NG \cite{11}. The study of Common vulnerability and exposures (CVEs) \cite{8} shows that there exist many vulnerabilities that allow unauthenticated access to the webcams these days. After capturing the victim’s audio from his home, the attacker can use the webcam to replay commands to a Google Home device in the absence of victim. This demonstrates a traditional replay attack with only one (1) point of replay. The webcam in this scenario could also be a baby monitor or other compromised VCD. We have verified through experiments that Google Home devices only authenticate the user based on the wake word “Hey Google”. As long as the audio recording of the user saying, “Hey Google” in this scenario is clear enough then an attacker can replay the wake phrase and insert any subsequent command i.e. “Open Garage Door”. 
\subsubsection{Scenario 2: Drop-In Replay}
 In the scenario 2 shown in Figure 2(b), we describe a situation where an attacker obtains an original recording of a victim (this can be accomplished for example by the attacker recording a phone call of the user). The attacker intends to replay the audio from his home to the victim’s home. The Amazon Echo devices offer a feature that allows audio conferencing between these devices. This feature a.k.a “Drop-In” works even between different homes and Echo devices owned by different people as long as their contact list permissions are set to allow that contact to “Drop-In”. When using the Drop-In mode, the receiving Amazon Echo device plays a small chime and changes the device’s light ring to green, thus enabling the conference mode. The presence of the recipient of the conference call is not required as the conference mode is enabled without any additional verification. If another VCD is nearby then commands can be replayed through the audio conference.
\\If an attacker is able to get himself, for example, on Bob’s contact list and allow himself the Drop-In permission then the attacker could start the audio conference between Amazon Echo VCDs at any time. The actions that the attacker would take are asking their Amazon Echo “Drop in on Bob”. The attacker could then replay the command “Hey Google, open the garage door”. Although this scenario requires that Bob’s Amazon Alexa contact list should be exploited by the attacker gaining access to Bob’s smartphone however, this scenario is plausible as a smartphone can be accessed by a trusted individual such as a friend, co-worker or child. 
\\This replay attack scenario demonstrates as proof-of-concept that VCDs in the home are vulnerable to replay attacks as long as the victim’s audio can be played in front of VCD in the home. This scenario is an example of a multi-hop replay attack as the original audio is replayed once to an Amazon Echo and then replayed again from the victim’s Amazon Echo device. We observed during the multi-hop replay scenarios that signal degradation due to multiple replays are unable to cause any problem as long as the playback audio is audible. While this scenario used two Amazon Echo devices, they could be replaced with other devices capable of transmitting audio such as a smart phone App being used to transmit audio to a webcam in another home.
\subsubsection{Scenario 3: Drop-In Multi-Home Replay}
 In scenario 3 (Figure 2c), we describe a situation where an attacker has managed to exploit a vulnerability in a webcam remotely and is able to access the camera’s audio and video stream. The camera is conveniently located in the kitchen near an Amazon Echo device. The attacker can easily identify the presence of the victims at home and observe their interactions with the VCD through accessing the camera. The Amazon Echo device itself does not have any voice verification system to verify the authenticity of any person. The attacker can cause chaos at the victim’s home during his absence by issuing commands to change the thermostat settings, turning on and off the lights, opening the garage door, etc. from the webcam. Every IoT device in the home that is connected to the Amazon Echo is now accessible to the attacker.
\\ The Amazon Echo “Drop-In” feature causes an additional threat in this situation. If the homeowner has allowed the Drop-In mode between friends, family members or even the workplace then the attacker will also have access to start an audio conference between other Amazon Echo devices. At this point the attacker could Drop-In to another Amazon Alexa located in another family members home. If there happens to be another VCD nearby, the attacker can then attempt to control IoT devices in the second home as well. In this scenario, the attacker can easily control another home remotely through the “Drop-In” feature.
\\ While we proposed few scenarios, it can be assumed that the device with the weakest security will be exploited. Many VCDs such as webcams are known to have vulnerabilities that expose their credentials or audio streams due to the nature of being mass produced cheaply in quick time. Once a VCD has been exploited then an attacker can have multiple options from listening and collecting audio to replaying audio or cloned voice. 
\begin{figure}
  \includegraphics[width=1\textwidth]{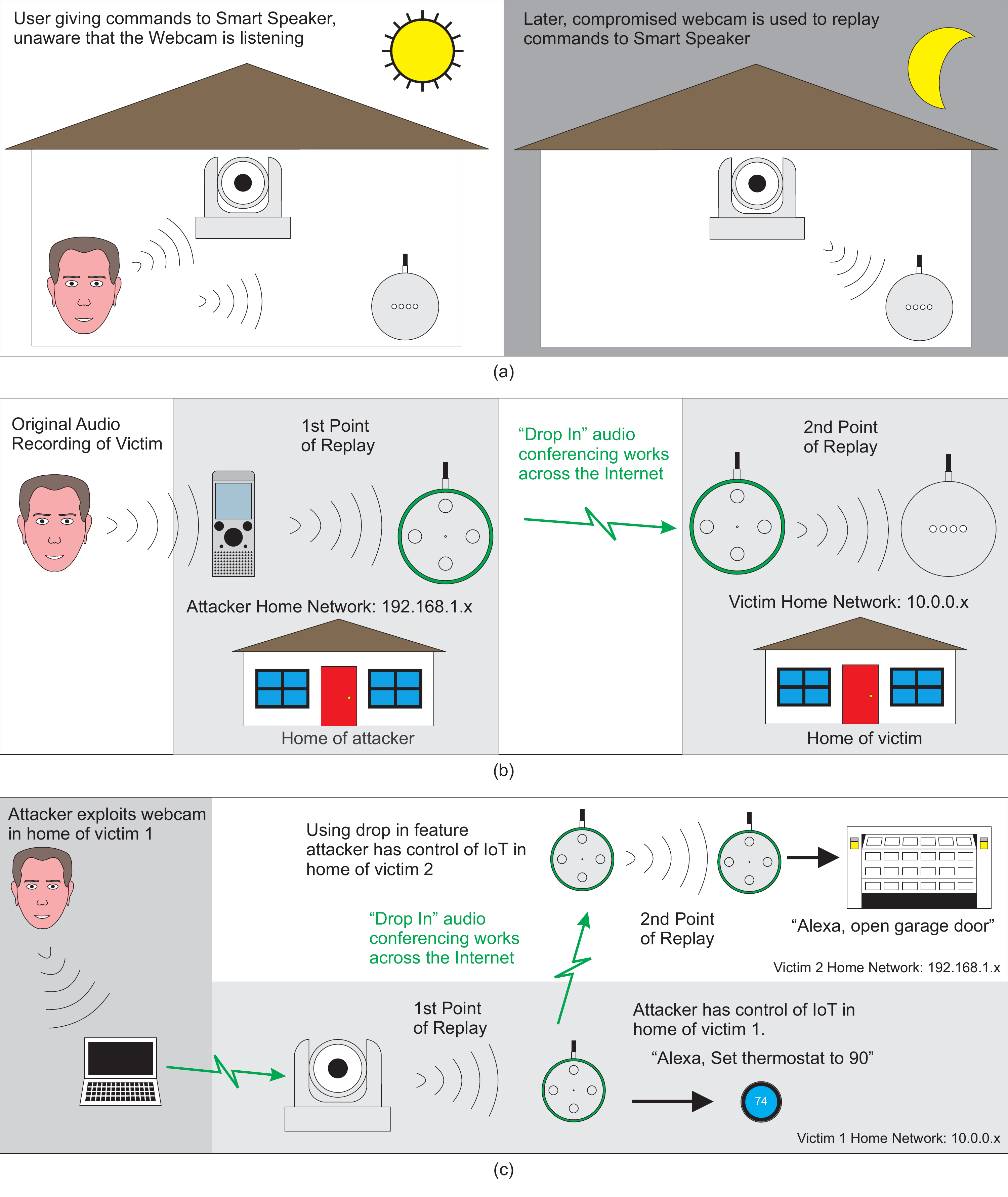}
\caption{Three replay attack scenarios. a) Webcam Replay.  b) Drop-In Replay. c) Drop-In Multi-Home Replay.}
\label{fig:2}       % Give a unique label
\end{figure}
\section{Dataset}
 This paper presents a unique voice spoofing detection corpus (VSDC) consisting of the bonafide, 1st-order and 2nd-order replay recordings by setting up different scenarios of chained VCDs. This multi-hop replay feature in our corpus can be used to evaluate the performance of different replay anti-spoofing algorithms in terms of classification among the bonafide, first-order and second-order replay attacks. More specifically, our dataset can effectively be used to investigate the performance of audio replay spoofing detection algorithms under diverse recording and playback environments, configurations, and devices. Additionally, our proposed VSDC can also be used to evaluate the performance of speaker verification systems as our corpus includes the audio samples of fifteen different speakers. It must be noted that all audio samples in our dataset are of six (6) seconds in duration to ensure uniformity when using the bonafide and spoof samples to train different machine-learning classifiers. 
\subsection{Definitions and Data Collection Strategy}
 As this paper discusses the idea of multiple points/order of replays, we need to define the terminology used to specify the given point of replay. We will refer to the bonafide recording of a person giving voice commands to a VCD as the zero point of replay or (0PR). When the original recording is replayed from an audio speaker, we will refer to the output audio as the first point of replay or (1PR). Similarly, when the 1PR audio is replayed through a chained VCD then we will refer to that output audio as the second point of replay or (2PR).
\\ Shown in Figure 3 is the process of capturing audio to create the dataset consisting of 0PR, 1PR and 2PR. The bonafide phrases (0PR files) can be captured on any recording device. The 0PR files are then copied to a PC for generating the replays. The PC replays the 0PR file to an audio speaker creating the 1PR audio. VCDs are setup in an audio conference mode so that the audio played at 1PR is replayed by the VCD at 2PR. The PC used for creating the data sample is simultaneously replaying the 0PR file while capturing the resulting 1PR and 2PR audio. The USB sound card connected to the PC in Figure 3 can be replaced by the onboard sound card of the PC or with a sound interface box.
\begin{figure}
  \includegraphics[width=1\textwidth]{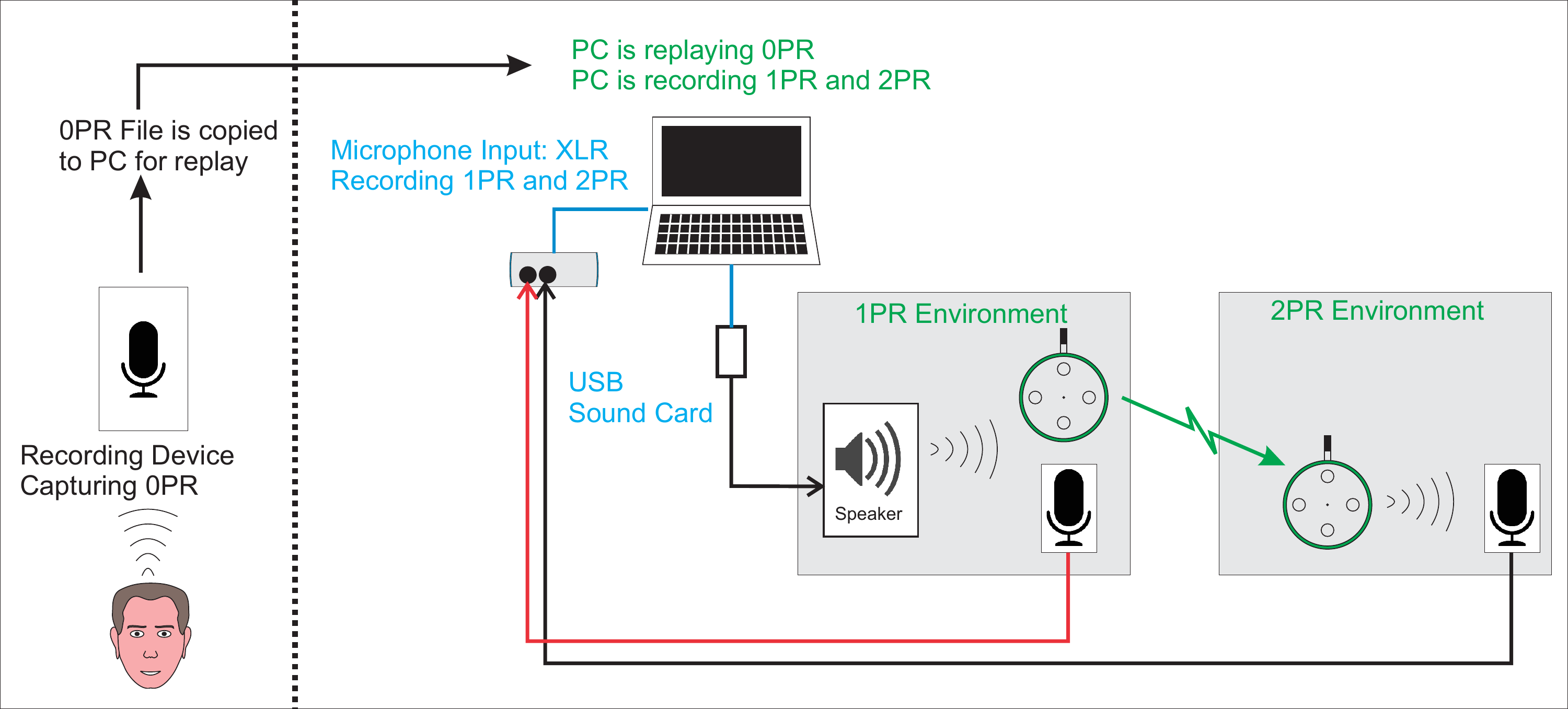}
\caption{Dataset creation configuration.}
\label{fig:3}       % Give a unique label
\end{figure}
\\ For the data collection, we used the Audacity tool \cite{9} to simultaneously play the 0PR audio while capturing the resulting 1PR and 2PR audio. The audacity tool can play from one audio track while simultaneously records audio on other tracks. To capture replays, a scenario is setup like two Amazon Echo devices in the conference mode to create a chain of VCDs. Audacity is setup with the bonafide (0PR) recording on track 1. The PC plays the audio to a connected audio speaker, this output audio is recorded on track 2 by Audacity and becomes the 1PR recording. At the same time, VCD replays the audio to the next device in the chain. The resulting output is recorded on track 3 by Audacity and becomes the 2PR recording. Using this method, we captured the 1PR and 2PR replays in real time. It was necessary to maintain proper isolation between the 1PR and 2PR environments to ensure that each respective microphone would not receive the same sound as the other (e.g. the 2PR microphone would not overhear the sound coming from the audio speaker used in the 1PR environment). Audacity is then used to trim the samples into 6 second duration. All data samples at 0PR, 1PR and 2PR are exported as separate files.
\subsection{Voice Commands and Recording Subjects}
 Shown in Table 1, forty-two (42) different phrases consisting of typical commands given to VCDs are used in creating the bonafide recordings. All commands start with the activation phrase “Hey Google”, “Computer” or “Alexa”. Some of the voice commands used for recordings are as follows, “Ok Google, turn on the kitchen light” or “Computer, turn off living room light”. The phrasing of giving a command using the activation word “Computer” reflects that replay attacks are not a vendor specific issue. These genuine utterances form the basis of our bonafide recordings.  A total of fifteen (15) human speakers, ages 18-60 years old, participated in data collection. Out of fifteen (15) human speakers, ten (10) are male, and five (5) are female. Some of the human speakers are not native English speakers. Each human speaker recorded the original file by repeating a given set of phrases typical of commands given to VCDs. Some of the volunteers recorded these original phrase sets multiple times using different microphones in diverse environments. In total 198 different 0PR source sets, consisting of at least 9 phrases, are created resulting in 1,926 0PR source phrases being spoken.

\begin{table}
\caption{Phrases used for 0PR samples.}
\label{tab:1}       % Give a unique label
\begin{tabular}{lll}
\hline\noalign{\smallskip}
Phrases  \\
\noalign{\smallskip}\hline\noalign{\smallskip}
Computer, turn on office lamp & Hey Google, turn off bedroom lamp \\
Computer, turn off office lamp & Hey Google, turn on living room light\\
Computer, turn on kitchen lights &	Hey Google, turn off living room light \\
Computer, turn off kitchen lights &	Hey Google, who am I \\
Computer, turn on bedroom lamp & Hey Google, give me an easter egg \\
Computer, turn off bedroom lamp	& Hey Google, good morning \\
Computer, turn on living room light	& Hey Google, tell me a joke \\
Computer, turn off living room light & Hey Google, beam me up \\
Computer, who am I	& Hey Google, set phasers to kill \\
Hey Google, turn on office lamp &	Hey Google, tea, earl grey, hot \\
Hey Google, turn off office lamp &	Hey Google, my name is Inigo Montoya\\
Hey Google, turn on kitchen lights & Hey Google, I want the truth \\
Hey Google, turn off kitchen lights	& Hey Google, turn on bedroom lamp\\
Alexa Turn, on office lamp & Alexa, my name is Inigo Montoya\\
Alexa Turn, off office lamp	& Alexa, I want the truth\\
Alexa, turn on kitchen lights & Alexa,turn off kitchen lights	\\
Alexa, turn on bedroom lamp &	Alexa, turn off bedroom lamp \\
Alexa, turn on living room light & Alexa, turn off living room light\\
Alexa, give me an Easter egg & Alexa, good morning\\
Alexa, tell me a joke &	Alexa, beam me up\\
Alexa, set phasers to kill &	Alexa, tea, earl grey, hot\\

\noalign{\smallskip}\hline
\end{tabular}
\end{table}
\subsection{0PR Environments}
 The 0PR environment is the area where the bonafide sound samples are recorded. The voice spoofing detection corpus (VSDC) includes the samples recorded at ten (10) different unique environments (for 0PR) that contain different amounts of ambient noise. We recorded the samples in different environments to ensure the diversity. The environments considered “noisy”, are the Computer Lab with Music, Car Off with light rain, Car on with Light Rain, Cafeteria, and University CourtYard. The Computer Lab with Music environment contained ambient noise from loud music. Both the Car on with Light Rain and Car Off with Light Rain environments are deemed as noisy due to the consistent pattering of rain and surrounding cars. The Cafeteria environment contains the undertones from discourse. The University Courtyard environment contains rustling trees and background conversations. The environments classified as low noise are, the Office Desk, Kitchen Table, Bedroom, and Computer Lab. All indoor environments contain less significant background noise from air circulation systems. 
\begin{figure}
  \includegraphics[width=1\textwidth]{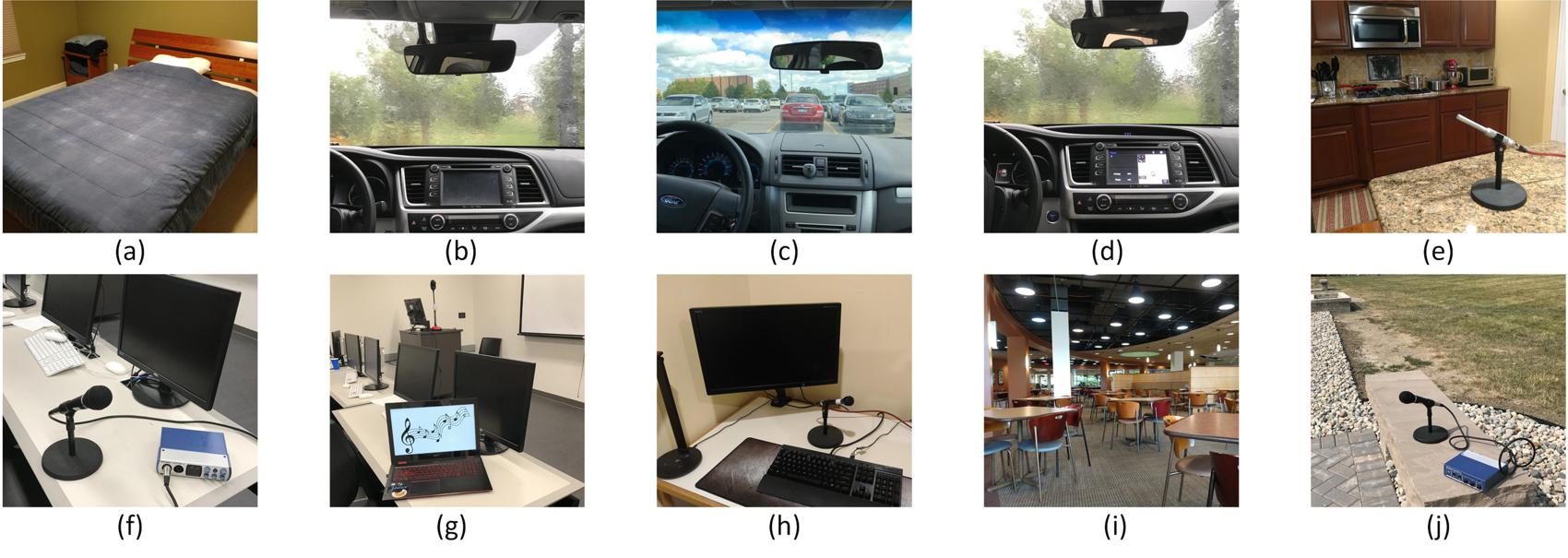}
\caption{Environments used for 0PR Recordings. (a) Bedroom, (b) Car Off Light Rain, (c) Car On, (d) Car On Light Rain, (e) Kitchen Table,
(f) Computer lab, (g) Computer Lab with Music, (h) Office Desk, (i) University Cafe, (j) University Courtyard.}
\label{fig:4}       % Give a unique label
\end{figure}
\subsection{Playback Environments}
 For playback environments, we used the mini audio booths, lab classrooms, and library study rooms to create the 1PR and 2PR replays. The brief description of each environment is given below.
\begin{figure}
  \includegraphics[width=1\textwidth]{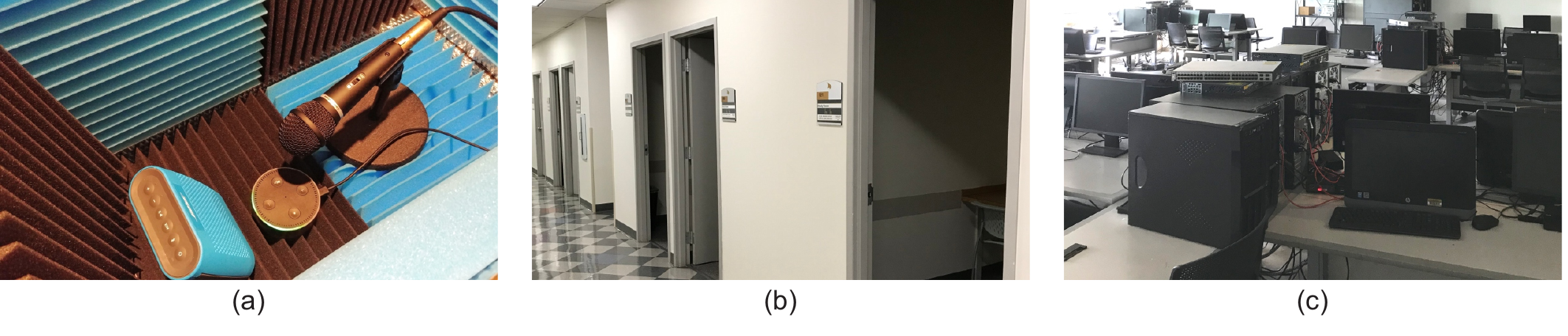}
\caption{Playback Environments.(a) Inside Mini recording booth, (b) Study Room, (c) Computer Lab.}
\label{fig:5}       % Give a unique label
\end{figure}
\paragraph{Mini Audio Booths} As shown in Figure 5(a), we designed the mini audio booths to eliminate the background sounds, i.e. sound of computer fans and air conditioners. Producing replay recordings in the audio booths allowed us to analyze the audio signals without ambient noise. The mini audio booths are effective in reducing ambient environmental noises. We used the SPL meter to determine the sound pressure level that recorded 40dBA for quiet office and 35.8 dBA inside the mini audio booth. Further testing shows that a large fan running 40 inches away from the microphone would produce a sound level of 56.5 dBA and 40.8 dBA in the office and audio booth respectively. Moreover, running a vacuum cleaner produces a level of 70 dBA in the office and 42.5 dBA inside the audio booth. More specifically, we created two audio booths one each for 1PR recordings and 2PR recordings. Shown in Figure 5(a) is the setup at 1PR, where a Bluetooth speaker is replaying an original audio sample. The Amazon Echo Dot is set to the “Drop-In” audio-conferencing mode so that it can replay the audio to another Echo Dot in second audio booth.
\paragraph{Study Room} We selected two study rooms inside the library to design the 1PR and 2PR playback environments (Figure 5b). In the first study room, we setup the 1PR playback equipment that includes the laptop, one Echo dot device, microphone and connecting cables. In the second study room, we created the 2PR playback environments where we used an Echo dot device and a microphone connected with the cables. It should be mentioned that these study rooms contained the noise from the air circulation system and footsteps.
\paragraph{Computer Lab} The computer lab playback environment used two computer labs adjacent to each other. One computer lab (Figure 5c) is used for conducting the 1PR replay, where we placed a speaker, microphones and the Echo Dot in “Drop-In” mode. In the next computer lab, we arranged the other Echo Dot along with the necessary microphone to capture the 2PR playback. The computer labs contain the noise of air circulation systems only.
\subsection{Equipment used for recording and playback}
 For recording the bonafide 0PR source files and the 1PR and 2PR replays, we used several combinations of microphones and microphone interface devices. More specifically, we used 22 distinct microphones for audio recordings. Shown in Table 2 are the combinations of external microphones and their microphone interface device. Devices with internal microphones are mentioned by the device name. Several professional grade microphones that use an XLR connection are used for recording and playback. These microphones are connected to the PC using an audio interface box. The professional microphones connected by XLR cables are highlighted in green in Table 2. An external USB microphone is used for making 0PR recordings. This type of microphone can be characterized as a medium quality microphone and is highlighted in yellow in Table 2. We also used various internal microphones of the laptops and cell phones. Internal microphones can be characterized as lower quality microphones and are highlighted in red in Table 2.
\begin{table}
\caption{List of all recording microphones and sound cards used.}
\label{tab:2}       % Give a unique label
\begin{center}
\begin{tabular}{l}
\hline\noalign{\smallskip}
\textbf{0PR Recording Configuration}\\
\noalign{\smallskip}\hline\noalign{\smallskip}
\rowcolor{green}
Audio-Technica ST95MKII $|$ Zoom R16	\\
\rowcolor{green}
Audio-Technica ST95MKII $|$ Presonus Studio 24	\\
\rowcolor{green}
Shure SM58 $|$ Zoom R16\\
\rowcolor{green}
Shure SM58 $|$ Presonus Studio 24\\
\rowcolor{green}
Behringer ECM8000 $|$ Zoom R16\\
\rowcolor{green}
Electro-Voice 635A/B $|$ Zoom R16\\
\rowcolor{yellow}
Blue Yeti $|$ Mac Book Pro 2018\\
\rowcolor{pink}
MacBook Pro 2018 (Internal Microphone)\\
\rowcolor{pink}
Acer (Internal Microphone)\\
\rowcolor{pink}
Samsung Galaxy S7 (Internal Microphone)\\
\rowcolor{pink}
IPhone 5S (Internal Microphone)\\
\rowcolor{pink}
(7) Android Phones\\
\hline\noalign{\smallskip}
\textbf{1PR and 2PR Recording Configuration}
\\
\noalign{\smallskip}\hline\noalign{\smallskip}
\rowcolor{green}
Audio-Technica ST95MKII $|$ Zoom R16\\
\rowcolor{green}
Audio-Technica ST95MKII $|$ Presonus Studio 24\\
\rowcolor{green}
Shure SM58 $|$ Zoom R16\\
\rowcolor{green}
Shure SM58 $|$ Presonus Studio 24\\
\rowcolor{green}
Behringer ECM8000 $|$ Zoom R16\\
\rowcolor{green}
\noalign{\smallskip}\hline
\end{tabular}

~\\ \textsuperscript{x}Formatted as [Microphone] $|$ [soundcard] or [device]. Devices contains an internal microphone and soundcard.
\end{center}
\end{table}

 Shown in Table 3 are the fourteen (14) 1PR playback configurations used in the proposed VSDC. The composition of configurations consists of a speaker, amplifier, and a sound card. We used a variety of speakers ranging from low to high quality. Devices such as laptops and cell phones that contain built in speakers represent the low quality, whereas, those using an external speaker either connected via Bluetooth or aux cable are considered devices of medium quality. Finally, the higher quality speakers are considered to be the external speakers with the manufacturer's specifications reporting that they produce a sound frequency response near the full range of human hearing of about 80hz - 20,000hz.
\begin{table}
\caption{List of all Playback configurations used in playing back audio in the 1PR recording environment.}
\label{tab:3}       % Give a unique label
\begin{tabular}{l}
\hline\noalign{\smallskip}
\rowcolor{white}
\textbf{1PR Playback Configuration}
\\
\noalign{\smallskip}\hline\noalign{\smallskip}
\rowcolor{green}
Polk R150 Speaker $|$ Yamaha HTR-5840 $|$ ZOOM R16\\
\rowcolor{green}
Polk R150 Speaker $|$ Yamaha HTR-5840 $|$ Asus GL504GM-DS74\\
\rowcolor{green}
Polk R150 Speaker $|$ Yamaha HTR-5840 $|$ USB Audio Card Ugreen 30521\\
\rowcolor{green}
Polk R150 Speaker $|$ Fisher 143 $|$ USB Audio Card Ugreen 30521\\
\rowcolor{green}
Bose 141 Speaker $|$ Yamaha HTR-5840 $|$ USB Audio Card Ugreen 30521\\
\rowcolor{green}
Presonus Eris E5 $|$ ZOOM R16\\
\rowcolor{green}
Presonus Eris E5 $|$ USB Audio Card Ugreen 30521\\
\rowcolor{yellow}
Bose Soundlink 415859 $|$ Asus GL504GM-DS74 (Wired)\\
\rowcolor{yellow}
 Bose Soundlink 415859 $|$ Asus GL504GM-DS74 (BlueTooth)\\
\rowcolor{yellow}
SBT6050R $|$ Asus GL504GM-DS74(Wired)\\
\rowcolor{yellow}
SBT6050R $|$ Asus GL504GM-DS74(BlueTooth)\\
\rowcolor{pink}
MacBook Pro 2018 (Internal Speaker)\\
\rowcolor{pink}
Acer Nitro Spin 5 (Internal Speaker)\\
\rowcolor{pink}
Acer Aspire E5-574G (Internal Speaker)\\
\hline\noalign{\smallskip}
\end{tabular}

\end{table}

 Eight (08) different configurations of devices are used to transmit the 1PR audio having the corresponding 2PR recordings as shown in Table 4. The device configurations vary from Amazon Echo’s using the “Drop-In” audio conferencing feature to laptops and tablets connected using the Google Meet.
\begin{table}
\caption{VCD, 1PR to 2PR Replay Configuration.}
\label{tab:4}       % Give a unique label
\begin{center}
\begin{tabular}{l l l}
\hline\noalign{\smallskip}
Source & Target & Connection Method\\
\noalign{\smallskip}\hline\noalign{\smallskip}
Echo Dot 2 & Echo Dot 2 & Amazon Drop-In\\
Echo Dot 2 & Echo Dot 3 & Amazon Drop-In\\
Echo Dot 3 & Echo Dot 2 & Amazon Drop-In\\
Echo Dot 3 & Echo Dot 3 & Amazon Drop-In\\
Echo Dot 3 & Echo Plus Gen 2 & Amazon Drop-In\\
Echo Dot 3 & Echo Input & Amazon Drop-In\\
LG G6 &	Asus Tablet	& Google-Meet\\
Laptop	& Laptop &	Google-Meet\\
\noalign{\smallskip}\hline
\end{tabular}
\end{center}
\end{table}
\\ The Amazon Drop-In audio conferencing offers easy transmission of audio from one location to another using the VCDs. We used several different combinations of Amazon Echo devices to test their audio quality. All of the Amazon devices even the previous Generation 2 Echo Dot with smaller speakers are able to replay commands to another VCD with acceptable results. All Amazon Echo devices contain audio out jacks which allow them to be connected to external speakers for improved quality. We connected the external speakers for various sample sets to analyze the changes in the audio signal. We used a variety of external speakers ranging from small battery powered external speakers, home theater speakers to studio monitor speakers.
\\ Google-Meet is used as an audio-conferencing service to transmit the 1PR audio to 2PR. While Google-Meet itself is used on laptops and tablets instead of the VCDs, still it is a useful way to replay audio to other VCDs. Laptops, tablets and phones can easily be left at other locations with the intention of replaying audio to VCDs at a later time. We used the laptops and Google-Meet to ensure the use of high-quality microphones. The audio quality of the replays is limited to the quality of Amazon Echo’s microphone in the configurations where we used the Amazon Echo devices. It can be expected that competition amongst the VCD manufacturers will continue to improve their audio capabilities likely to the point that products will become available for the audiophile market. Therefore, we conclude that it is worthwhile to study the audio characteristics of replay attacks on all types of audio equipment.
\subsection{Data Availability}
 The dataset is organized into different folders, where each folder has all the recordings (0PR, 1PR and 2PR) of unique speaker. Each speaker folder contains three sub-folders including the audio samples of 0PR, 1PR and 2PR. The naming convention of the file specifies the sample number, point of replay, speaker, environment, microphone at 0PR and configuration number as shown in Table 5. The proposed voice spoofing detection corpus is available at \cite{10} for research purposes.
\begin{table}
\caption{VCD, 1PR to 2PR Replay Configuration.}
\label{tab:5}       % Give a unique label
\begin{center}
\begin{tabular}{l l}
\hline\noalign{\smallskip}
\textbf{Sample set} & 11\\ 
\textbf{point of Replay (PR)} & 1PR \\
\textbf{Speaker} & S2\\
\textbf{Environment} & E6\\
\textbf{Microphone used for recording} & M7\\
\textbf{Configuration Number} & C28\\
\textbf{Phrase} & 01\\

\noalign{\smallskip}\hline
\end{tabular}
\end{center}
\end{table}
\\ Sample Set: Indicates the original 0PR file the sample is based from.
\\ Point of Replay: Indicates at which point of replay this sample was created.
\\ Speaker: Human speaker/volunteer who recorded this sample.
\\ Environment: Recording Environment of the 0PR sample.
\\ Microphone used for recording: The microphone that was used for the 0PR recording.
\\ Configuration Number: The configuration setup that was used to make the 1PR and 2PR samples. Configuration is based on: Replay Speaker, Replay Device, 1PR to 2PR transmission device and method.
\\ Phrase: In each configuration the volunteer spoke at least 9 phrases. This number indicates which phrase it is out of all the phrases spoken for that sample
\\ The resulting filename looks like this: 11-1PR-S2-E6-M7-C8-01
\\ Shown in Table 6 are the number of samples collected at each point of replay. The word sample refers to each 6 second audio file that can have any phrase listed in Table 1. The 0PR samples are the original bonafide phrases used. The 0PR files are replayed multiple times with unique microphone and speaker configurations to create the 1PR and 2PR samples. 
\begin{table}
\caption{Number of Samples Spoken.}
\label{tab:6}       % Give a unique label
\begin{center}
\begin{tabular}{l l}
\hline\noalign{\smallskip}
\textbf{Sample Type} & \textbf{Number of Samples}\\ 
0PR	& 1926 \\
1PR	& 4923\\
2PR	& 4923\\
Total &	11772\\
\noalign{\smallskip}\hline
\end{tabular}
\end{center}
\end{table}
\section{Experiments and Results}
 To indicate the significance of our dataset, we performed different experiments to evaluate its performance against two well-known datasets: ASVspoof 2019 and ASVspoof 2017 \cite{2,3}. Since constant Q cepstral coefficients (CQCC) features and gaussian mixture model (GMM) classifier were used for evaluation of ASVspoof 2017 and 2019, we used the same method for fair comparison. We used the CQCC-GMM model for performance evaluation since it is recommended as a baseline model by the ASVspoof. We used only ASVspoof baseline methods as the experiments were designed to measure the significance of our dataset to compare performance with ASVspoof datasets. For evaluation, ASVspoof 2017 used the metric of equal error rate (EER) while ASVspoof 2019 employed min-tDCF (tandem-decision cost function) along-with the EER. Therefore, we used the EER metric for ASVspoof 2017 and, EER and min-tDCF metrics \cite{4} for ASVspoof 2019 to evaluate the performance. 
\subsection{Experiment-1: Training on ASVspoof dataset}
 We designed one experiment to investigate the capability of the ASV anti-spoofing system for replay detection in more diverse conditions. For this purpose, we trained the ASVspoof 2017 baseline CQCC-GMM model on the training samples of ASVspoof 2017 database version-2 \cite{2} and tested it on both the ASVspoof 2017 dataset (development + evaluation) and our voice spoof detection corpus (VSDC) \cite{10}. ASV 2017 baseline model provided an EER of 29.95\% on the evaluation set and 12.08\% on the development set of ASVspoof 2017, however the performance of baseline method degrades significantly on VSDC resulting in EER of 47.68\%. Similarly, we trained the ASVspoof 2019 baseline CQCC-GMM model using the training samples of ASVspoof 2019 dataset \cite{3} and evaluated on both the ASVspoof 2019 dataset (development + evaluation) and VSDC. We obtained an EER and min-tDCF of 22.66\% and 0.414 on ASV development dataset, and 23.16\% and 0.624 on ASV evaluation dataset. Moreover, we achieved an EER and min-tDCF of 30.65\% and 0.636 respectively on VSDC. The results of EER are shown in Figure 6, whereas Figure 7 shows min-tDCF values. 
\begin{figure}
 \includegraphics[width=1\textwidth]{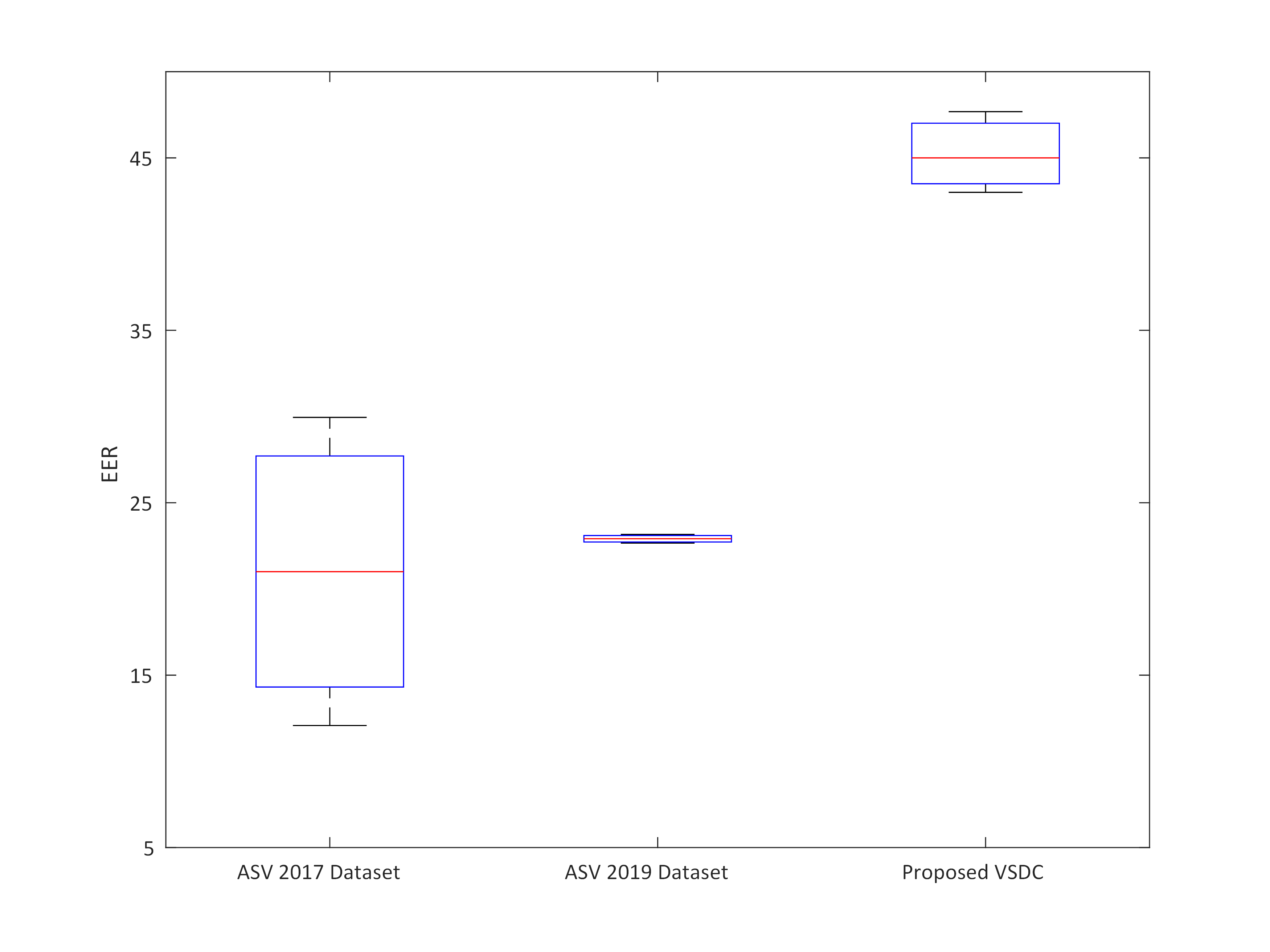}
\caption{Error Rate of ASVspoof 2017 and 2019 baseline methods on ASVspoof and Proposed VSDC Datasets.}
\label{fig:6}       % Give a unique label
\end{figure}
\begin{figure}
 \includegraphics[width=1\textwidth]{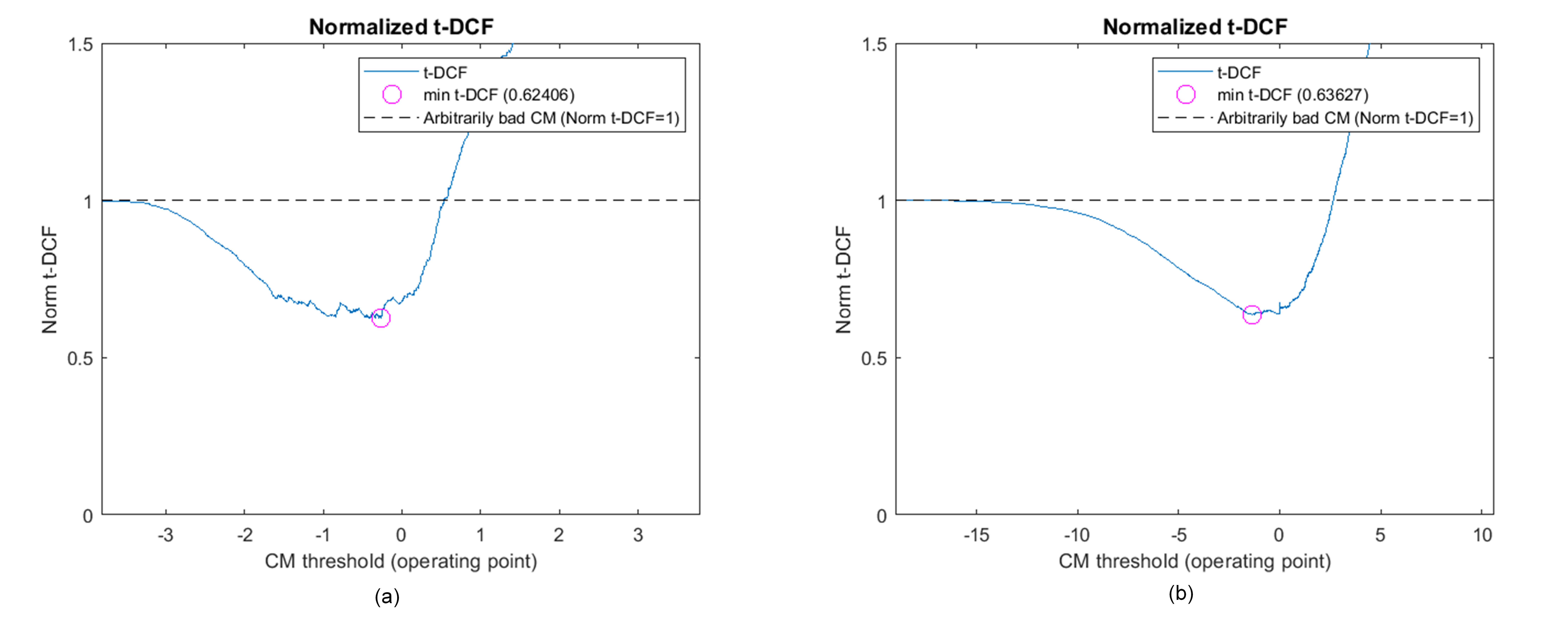}
\caption{tDCF value for ASV 2019 baseline method on (a) ASVspoof 2019 dataset, and (b) Proposed VSDC.}
\label{fig:7}       % Give a unique label
\end{figure}
\\ As the ASVspoof baseline method is trained on the audio samples collected in the indoor environments with minimum distance between the speaker and microphone, therefore, we ensured to create a diverse spoofing dataset by recording and playback in different environments (ambient background noise to anechoic recording chamber), using different recordings and playback devices to maintain diversity in acoustics signatures, and recordings in multi-hop scenarios. The results presented in Figure 6 clearly illustrate that ASV counter-preventive (CP) model has dependency on the environment, recording and playback settings/scenarios which makes this model unable to accurately detect the replay attacks under diverse environment conditions, and recording and playback scenarios. 
\subsection{Experiment-2: Training on ASVspoof and VSDC}
 To test our hypothesis that training the anti-spoofing system with more diverse dataset containing the attributes mentioned previously can enhance the performance of the baseline model, we designed our second experiment to train the ASVspoof baseline model on the combined corpus comprising of both the ASVspoof and our VSDC samples. We performed this experiment in two stages. First, we trained the baseline model on the combined corpus of ASVspoof 2017 training and VSDC followed by testing on the ASVspoof 2017 evaluation dataset. We achieved an EER of 25.24\% on this combined corpus that is 4.71\% lesser than the EER obtained by the model trained only on the ASVspoof dataset. In the second stage, we trained the baseline model on the combined corpus of ASVspoof 2019 training dataset and VSDC and tested it only on the ASVspoof 2019 evaluation dataset. We obtained an EER of 15.79\% and 17.39\% for ASVspoof 2019 development and evaluation datasets as shown in Figure 8. These results are 6.87\% and 5.77\% lesser than the EER achieved on the model trained only on the ASVspoof 2019 dataset. In addition, min-tDCF of 0.326 and 0.347 is achieved for development and evaluation dataset as plotted in Figure 9. From the results (Figure 8 and 9) of this experiment we can conclude that training the anti-spoofing model with additional audio samples collected in more diverse settings and scenarios can enhance the performance of the anti-spoofing model. 
\begin{figure}
 \includegraphics[width=1\textwidth]{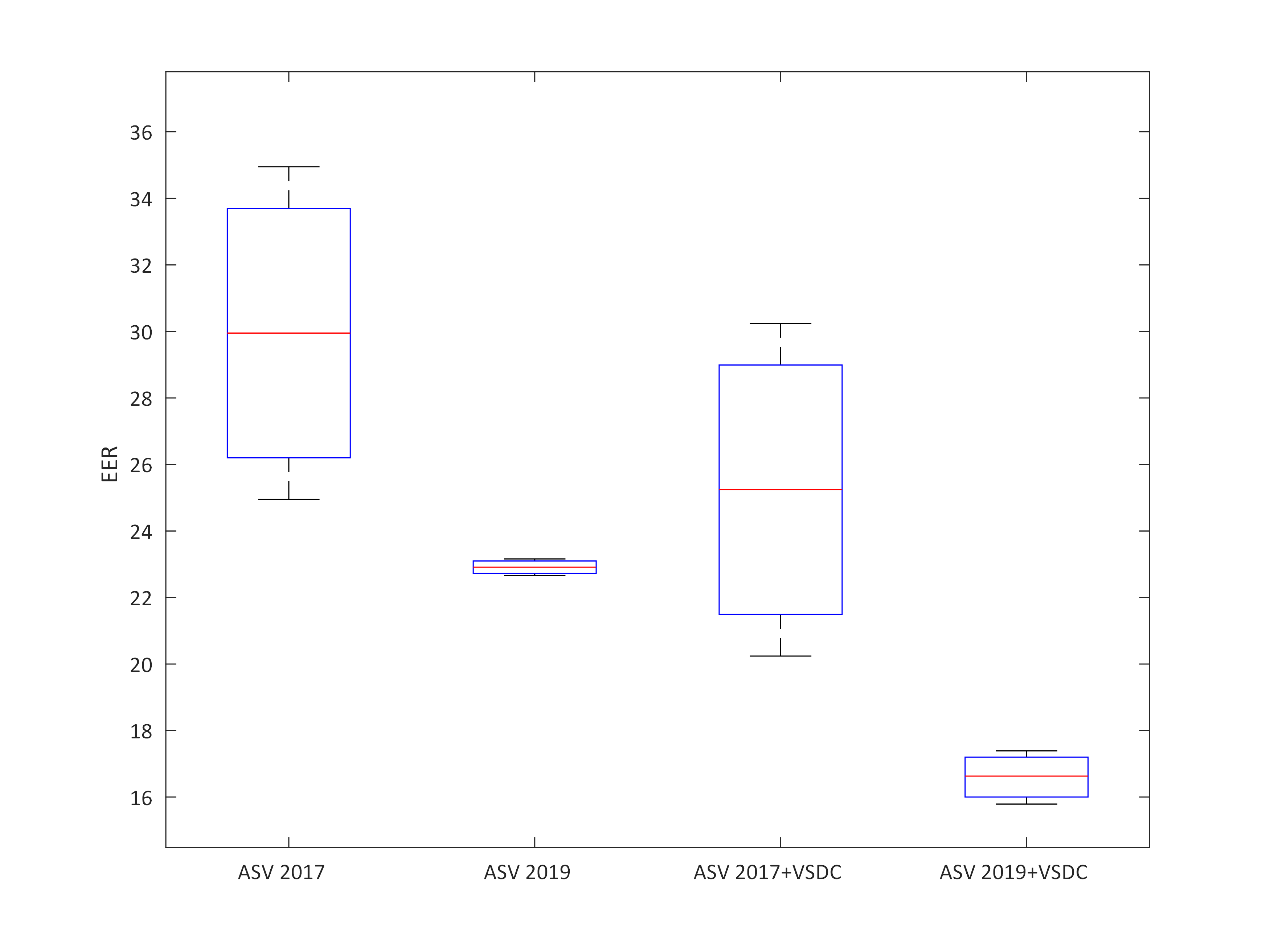}
\caption{Error Rate of ASVspoof 2017 and 2019 baseline methods trained on ASVspoof and ASVspoof-VSDC Datasets.}
\label{fig:8}       % Give a unique label
\end{figure}
\begin{figure}
 \includegraphics[width=1\textwidth]{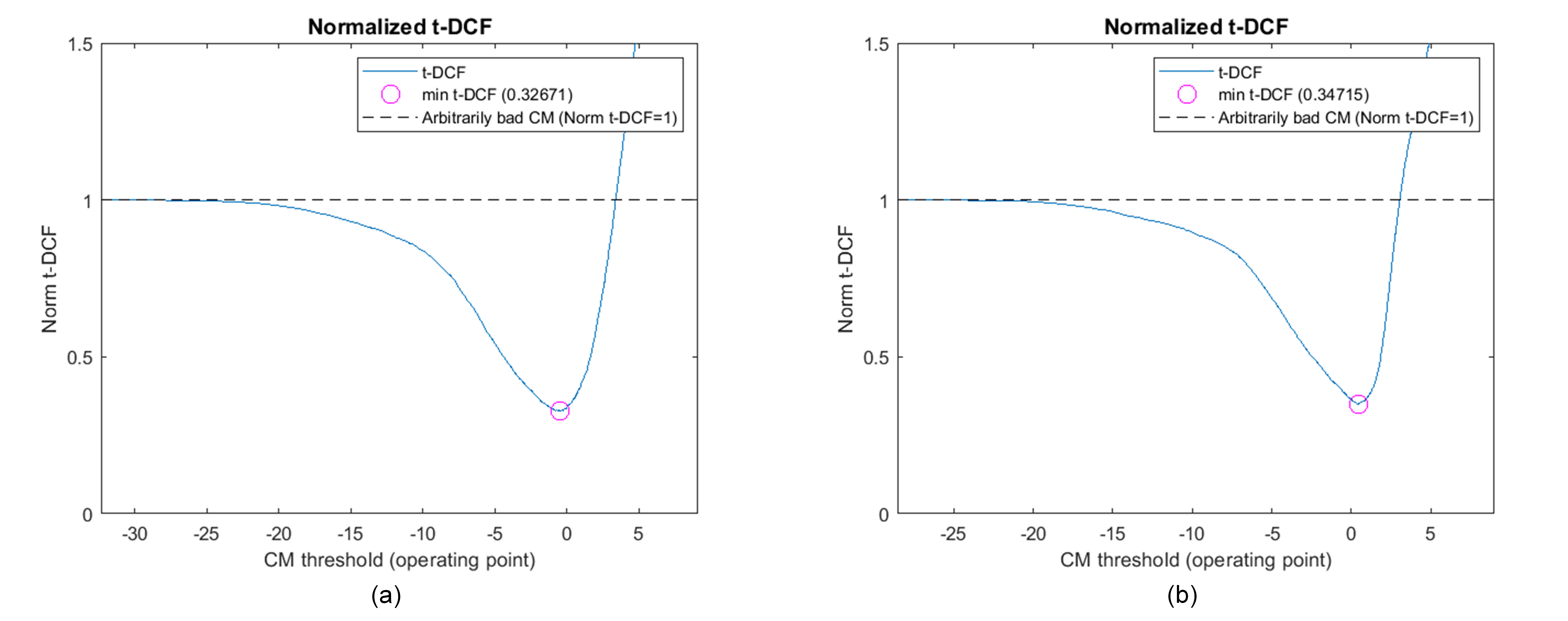}
\caption{tDCF value for ASVspoof 2019 baseline method on ASVspoof 2019 (a) development dataset, (b) evaluation dataset.}
\label{fig:9}       % Give a unique label
\end{figure}
\subsection{Experiment-3: Training on proposed VSDC for multi-order replay attacks}
 Our VSDC is unique to existing spoofing datasets in the way that we further categorize the spoofing samples into first-order and second-order replay attacks. We already demonstrated through experiments in our previous work [5] that ASV systems like Google Home and Amazon Alexa are even vulnerable to multi-hop scenarios (i.e. second-order replay attacks). Therefore, we argue that anti-spoofing systems must have the capability to accurately detect the second-order replay attacks as well besides the first-order replays. To test the effectiveness of our dataset in this perspective, we performed an experiment in two stages, one using the bonafide and first-order replay samples, and second using the bonafide and second-order replay samples.
\\In the first stage of this experiment, we evaluated the performance of ASV baseline method on our dataset using only the bonafide and first-order replay samples. For this purpose, we used 60\% samples to train the ASV baseline method where half of the samples belong to the bonafide and rest to first-order replays. ASV baseline method provided an average EER of 20.54\% on our dataset of bonafide and first-order replay samples as shown in Figure 10. In the second stage of this experiment, we evaluated the ASV baseline method using only the bonafide and second-order replay samples. We used similar experimental settings as adopted for first-stage experiment (60\% for training and 40\% for testing) and obtained an average EER of 10.74\% (Figure 10). The results of this experiment indicate that the first-order replays are more challenging to detect as compared to second-order replay attacks indicating the fact that the characteristics of playback devices have some correlation with the spoof samples. 
\begin{figure}
 \includegraphics[width=1\textwidth]{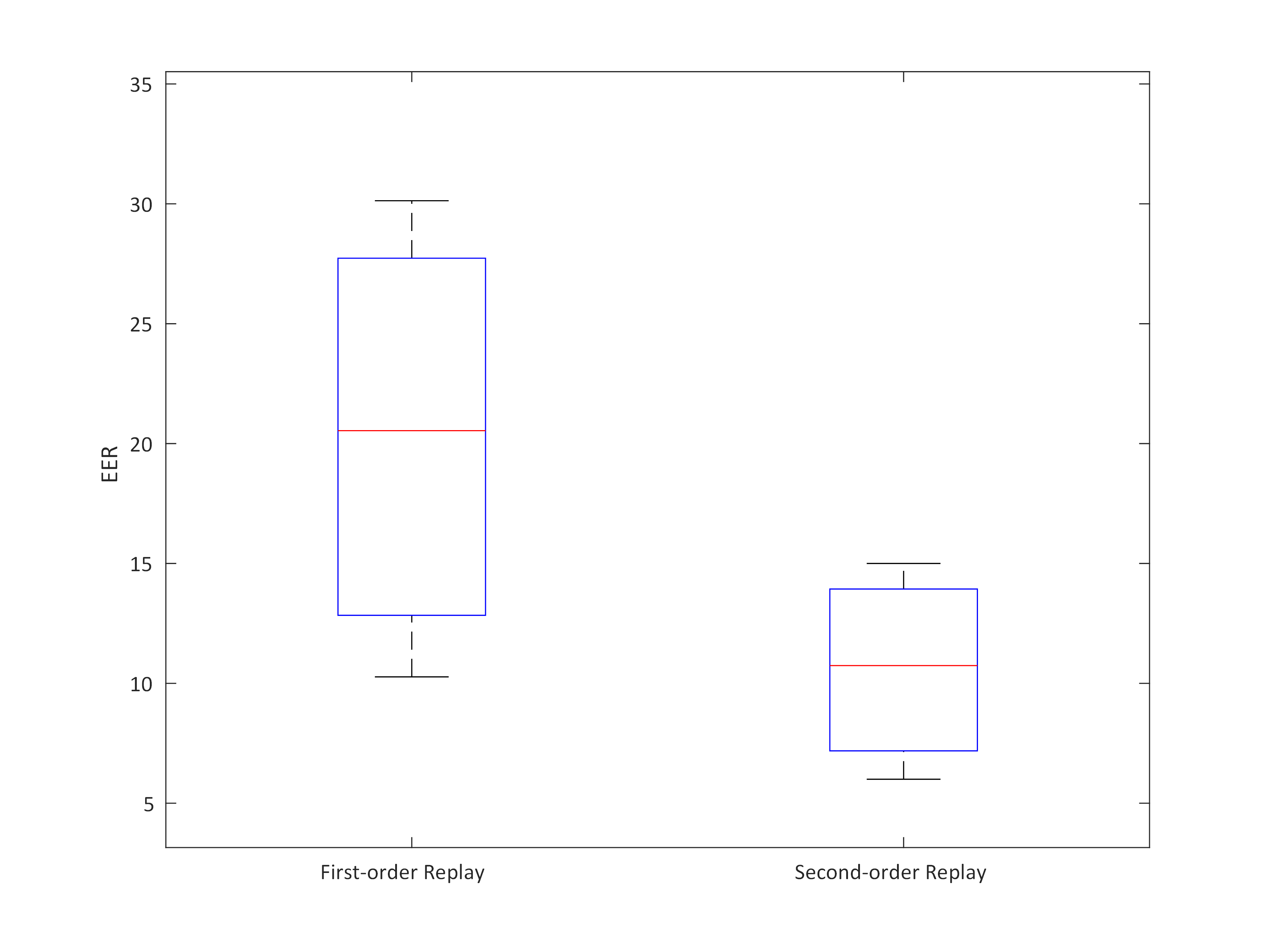}
\caption{Error Rate of ASVspoof baseline method on First-order and Second-order Replay Samples of the Proposed VSDC.}
\label{fig:10}       % Give a unique label
\end{figure}
\section{Conclusion}
In this paper, we presented an audio replay spoofing dataset with the motivation to address the limitations of existing replay spoofing datasets. In comparison of existing datasets like ASVspoof, our proposed dataset is diverse in terms of recording and playback scenarios (i.e. second-order replay attack, etc.), devices, environment, etc. According to the best of our knowledge, it is the first attempt to create an audio spoofing dataset for multi-hop scenarios. After performance evaluation of ASVspoof baseline method on the proposed dataset we witnessed a significant performance degradation due to more diverse scenarios of our dataset. On the other hand, the baseline model trained on different diverse scenarios eventually found to perform better than the previous case. It has also been observed after experimentation that the discrimination between the bonafide and first-order replay samples is more challenging as compared to second-order replay samples. Moreover, we conclude that the characteristics of playback devices must also be thoroughly investigated to identify the difference in features of the first-order replay and second-order replay spoofing samples. 
\\ The proposed dataset contributes to the existing spoofing dataset mainly through adding more diversity in playback scenarios (i.e. multi-hop replay attack), recording environments, and professional microphones. ASVspoof being a crowdsource dataset possess enough diversity in terms of recording devices, environments, and playback scenarios and can be used to evaluate anti-spoofing methods for both replay attacks and cloning. On the contrary, our dataset is specifically designed for multi-order replay spoofing detection and diverse enough to be effectively used to evaluate the replay anti-spoofing methods. As we already discussed the importance of multi-hop replay attacks in MoT environment, our dataset can reliably be used to evaluate the performance of anti-spoofing systems capable of handling the multi-order replays. However, ASVspoof dataset is unable to handle the multi-hop replay scenario that also indicates the potential gap in ASVspoof dataset. Therefore, we argue that the proposed VSDC can effectively be used to develop more robust anti-spoofing methods under diverse recording and playback configurations, replay scenarios including multi-order replay attacks, and recording and playback environments.
\\
\\
\textbf{ACKNOWLEDGEMENT} 
This work was supported by grant of National Science Foundation (NSF) of USA via Awards No. (1815724) and (1816019).

\end{document}